\documentclass[conference]{IEEEtran}
\usepackage{cite}
\usepackage{amsmath,amssymb}
\usepackage{graphicx}
\usepackage{textcomp}
\usepackage{url}
\usepackage{balance}

\begin{document}

\title{ClearTalk: Data-Driven Mandarin Articulation Analysis and Personalized Communication Support for Children with Dysarthria}

\author{\IEEEauthorblockN{Ben Xiaolu Huang}
\IEEEauthorblockA{\textit{High School Student}\\
\textit{Shanghai Foreign Language School Affiliated to SISU}\\
Shanghai, China\\
benhsfls@gmail.com}}

\maketitle

\begin{abstract}
Children with dysarthria may experience reduced intelligibility and difficulty expressing everyday needs. ClearTalk is a smartwatch-based system that converts Mandarin speech-recognition differences into interpretable articulatory profiles for personalized practice and staged communication. Target and recognized utterances are aligned at the syllable level, summarized across seven articulatory positions, and ranked by a count-sensitive asymmetric heuristic. For 13 dysarthric speakers from the Mandarin Dysarthria Speech Corpus, system syllable-level accuracy showed a Pearson correlation of 0.908 with expert intelligibility, with an R-squared value of 0.824 and a p-value below 0.001. The system-selected priority targets achieved an 86.7 percent overlap with the consensus of three blinded speech-language pathologists; across 210 target consonants, automated labels achieved 86.7 percent agreement, a Cohen's kappa of 0.76, and a macro-F1 score of 0.81. In a balanced single-session pilot with 15 children, mean pre/post gains were 14.4, 29.4, and 27.7 percentage points for mild, moderate, and severe groups, while communication-round completion rates were 81.3, 66.7, and 48.0 percent. These findings support the feasibility of using interpretable patterns of speech errors to guide wearable articulation practice and communication assistance.
\end{abstract}


\section{Introduction}
Dysarthria can reduce speech intelligibility and limit children's classroom, social, and everyday communication \cite{kent2003}. Motor-speech interventions address specific production patterns \cite{korkalainen2023,pennington2013}, while augmentative and alternative communication systems support message expression when speech is not consistently understood \cite{beukelman2020}. These functions are often separated: training tools emphasize repeated practice, whereas communication tools focus on message output without using articulation patterns to personalize later exercises.

ClearTalk connects these functions through a data-driven wearable workflow. The system aligns recognized and target Mandarin syllables, identifies recurring error types, summarizes initial-consonant performance across seven articulatory positions, and uses the resulting profile to guide practice generation. When complete recognition remains difficult, a staged interface supports message construction through scene and semantic selection.

This paper contributes: (1) an interpretable Mandarin syllable-analysis pipeline and seven-position articulation profile; (2) a heuristic mechanism for selecting personalized practice targets; (3) a staged object--action--need communication workflow; and (4) multi-level validation combining public-corpus correlation, blinded expert agreement, ablation, robustness analysis, and a severity-stratified child pilot.

\section{System and Methods}
\subsection{System Overview}
ClearTalk uses an ESP32-based smartwatch for audio capture, prompts, option selection, synthesized playback, and nonaudio session summaries. Cloud services provide Mandarin automatic speech recognition \cite{gao2023funasr}, constrained sentence and option generation through the Alibaba Cloud Model Studio Qwen-Plus model \cite{alibaba2026}, and speech synthesis. Under the tested network conditions, recognition--generation--synthesis requests typically completed in 3--5 s. The system contains two modules: personalized articulation practice and staged communication assistance.

\begin{figure*}[!t]
\centering
\includegraphics[width=\textwidth]{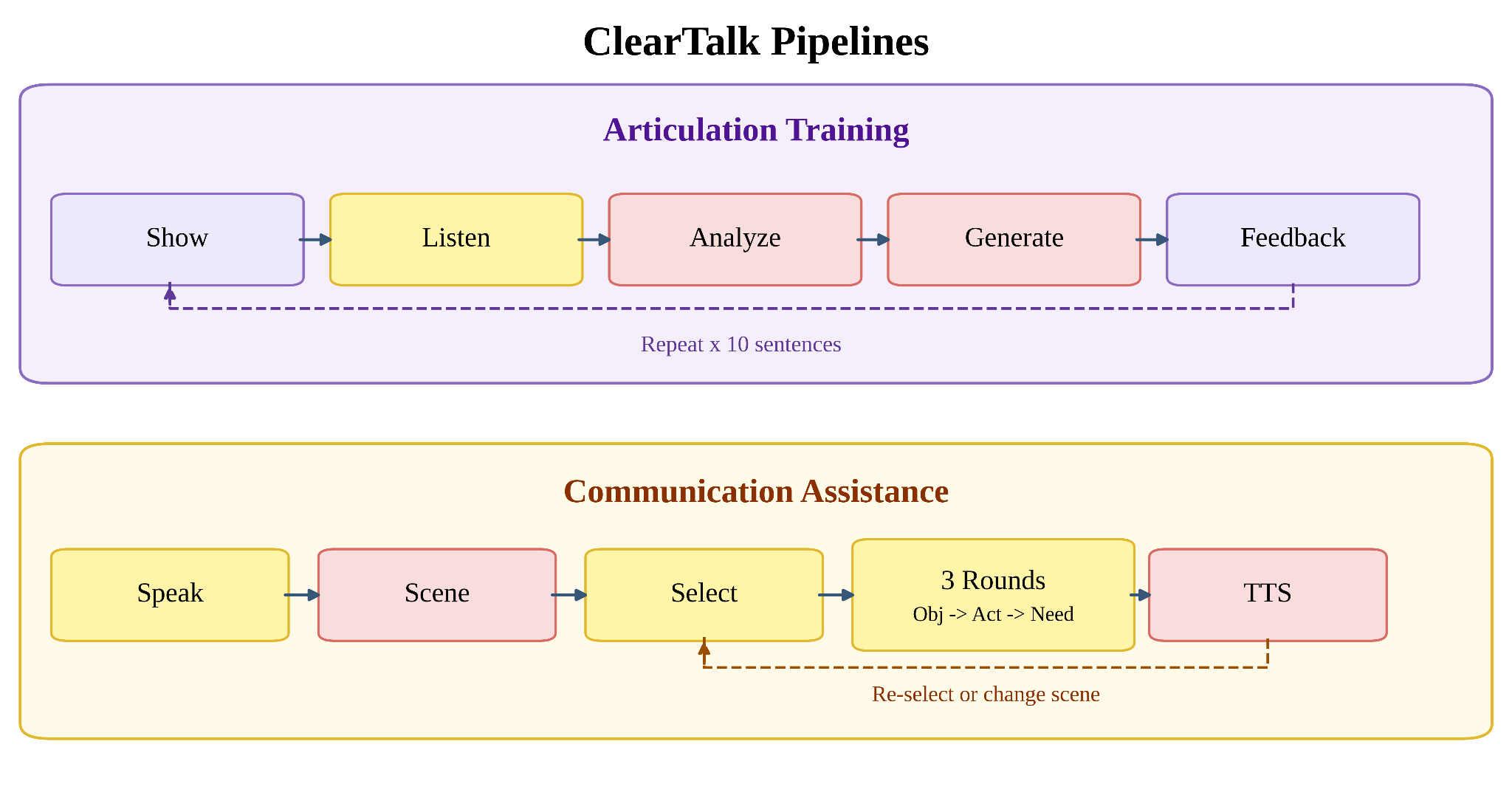}
\caption{High-level workflows of ClearTalk's articulation-practice and staged communication-assistance modules.}
\label{fig:pipeline}
\end{figure*}

The two module workflows are summarized in Fig.~\ref{fig:pipeline}.

\subsection{Syllable Alignment and Error Profiling}
The target sentence and automatic speech-recognition output are converted to pinyin syllable sequences and aligned at the syllable level. Each mismatch is categorized as an initial-consonant error, final-vowel error, tone error, deletion, or insertion. Initial consonants are mapped to seven Mandarin articulatory positions: bilabial (b/p/m), labiodental (f), alveolar (d/t/n/l), velar (g/k/h), alveolo-palatal (j/q/x), retroflex (zh/ch/sh/r), and alveolar/dental (z/c/s). For descriptive evaluation, position-level initial-consonant accuracy is calculated as the number of correctly aligned target consonants divided by the number of target consonants belonging to that position.

\subsection{Heuristic Personalization Score}
For real-time personalization, ClearTalk maintains a count-sensitive asymmetric heuristic score for each articulatory position. Each score begins at 75; a correctly recognized target syllable adds 5 points, while a target-anchored mismatch associated with that position subtracts 15 points. The score is clipped to the interval from 0 to 100:
\begin{equation}
H_k = \max\bigl(0,\min(100,75+5C_k-15E_k)\bigr),
\label{eq:heuristic}
\end{equation}
where $C_k$ and $E_k$ denote the numbers of correct and mismatched target syllables associated with position $k$. Insertion errors, which have no corresponding target initial consonant, contribute only to the global syllable-level error rate and are excluded from the position-specific penalty $E_k$. The three lowest-scoring positions are passed to Qwen-Plus \cite{alibaba2026}, which generates ten age-appropriate practice sentences targeting their consonant sets. The heuristic is used only to rank practice targets and is not interpreted as a clinical severity measure. For descriptive evaluation, position-level accuracy is reported separately as correctly aligned target consonants divided by all target consonants assigned to that position.

\subsection{Staged Communication Assistance}
The communication module reduces dependence on successful full-sentence recognition. After an initial speech attempt, the child selects a scene and completes three semantic rounds corresponding to object, action, and need. Each round presents four concise options, with refresh and undo functions available. The selected concepts are converted into concise, complete, and direct sentence alternatives, one of which can be played through text-to-speech synthesis. A child may revise the scene or selections before final playback.

\section{Evaluation}
\subsection{MDSC Dataset Evaluation}
The technical evaluation used the Mandarin Dysarthria Speech Corpus (MDSC) training subset \cite{gao2024mdsc}, which contains 13 dysarthric speakers and 21 control speakers. Each speaker recorded 355 non-wake-up utterances in addition to repeated wake-up words. For the primary speaker-level correlation and articulatory-position analyses, ten utterances were uniformly and randomly sampled from each speaker's 355 non-wake-up utterances using a fixed random seed of 42. The same cloud-processing pipeline and sampling procedure were applied to all speakers, resulting in 130 dysarthric and 210 control utterances, or 340 utterances in total.

For the 13 dysarthric speakers, system syllable-level accuracy was aggregated at the speaker level and compared with the five-annotator expert intelligibility scores supplied by MDSC. Expert scores were also used for descriptive grouping: mild (80 or above), moderate (60--79), and severe (below 60). Position-level initial-consonant accuracy was calculated for every speaker; group bars report means of speaker-level values, error bars show plus or minus one standard error of the mean, and dots show individual speakers.

\begin{figure*}[!t]
\centering
\includegraphics[width=\textwidth]{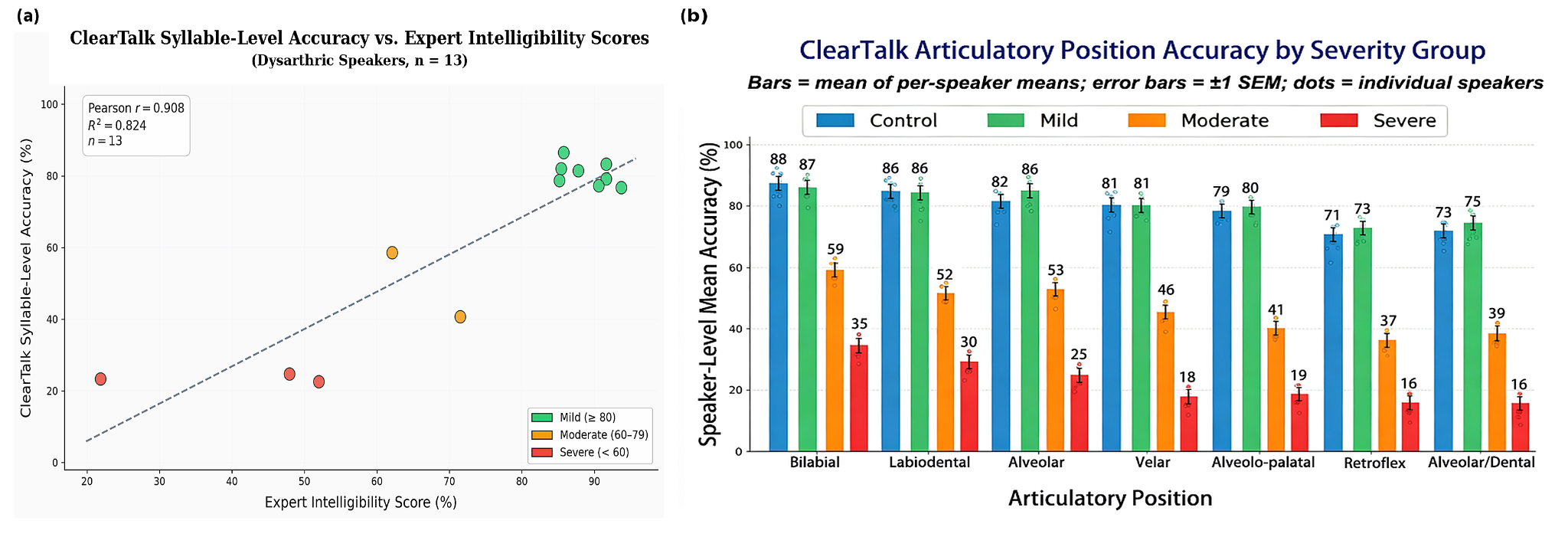}
\caption{MDSC evaluation. (a) ClearTalk speaker-level syllable accuracy versus expert intelligibility scores for 13 dysarthric speakers. (b) Descriptive initial-consonant accuracy across seven articulatory positions for control and dysarthria severity groups.}
\label{fig:mdsc}
\end{figure*}

The speaker-level and position-level analyses are summarized in Fig.~\ref{fig:mdsc}.

\subsection{Expert Alignment, Ablation, and Robustness Protocols}
Three certified speech-language pathologists (SLPs) independently reviewed all 75 baseline utterances (five per child) while blinded to system outputs. For each child, each rater selected and ranked three priority articulatory positions; consensus was formed by majority vote, with mean rank resolving ties. Top-3 overlap compared the heuristic with SLP consensus and with a normalized-accuracy baseline, $C_k/N_k$. MDSC sampling stability was evaluated across ten prespecified seeds, and local parameter sensitivity was tested over reward weights $\{4,5,6\}$ and error penalties $\{-12,-15,-18\}$.

For low-level validation, a stratified subset of 30 pre-test utterances (two per child) yielded 210 target initial consonants spanning all seven positions. Three SLPs, blinded to system outputs, independently labeled each target as correct, initial-consonant error, deletion, or uncertain. Majority vote formed the reference; unresolved cases were adjudicated by consensus, yielding final three-class labels. Evaluation used overall agreement, Cohen's kappa, macro-F1, and position-specific error recall.

\subsection{Child Pilot Protocol}
Fifteen children with dysarthria aged 6--12 years (mean age 9.3 years) participated in a balanced, severity-stratified single-session feasibility pilot. Three SLPs independently assessed baseline intelligibility from fixed reading materials and natural conversation while blinded to ClearTalk outputs. Mean ratings defined mild (80 percent or above), moderate (60--79 percent), and severe (below 60 percent) groups, with five children per group. Reliability of the three-rater mean was good under a two-way mixed-effects, absolute-agreement, average-measures model (ICC(A,3) = 0.88, 95 percent confidence interval 0.82--0.93, $p < 0.001$) \cite{koo2016}.

Each child completed five pre-test sentences, ten personalized practice sentences, and five post-test sentences. The pre/post sets were nonoverlapping researcher-constructed matched parallel sentence sets balanced in syllable length, syntactic complexity, and initial-consonant distribution across the seven positions. Practice sentences were generated after the pre-test and did not overlap with either assessment set; the post-test therefore examined immediate transfer to unpracticed but phonetically matched items.

Each child completed five daily scenarios---requesting a drink, asking for help, expressing feelings, describing pain, and greeting---each with object, action, and need selection rounds. This produced 15 rounds per child and 75 per severity group. A round counted as independent only when the intended option was selected without verbal prompting, option indication, or physical assistance. Each child--guardian dyad then rated six usability dimensions on a five-point scale.

\subsection{Ethics and Data Handling}
The school-based evaluation was conducted under project-advisor supervision. Written guardian consent and age-appropriate child assent were obtained; participation was voluntary and could stop at any time. The protocol did not undergo formal institutional ethics-board review, so findings are presented only as non-clinical feasibility evidence. Audio was processed transiently without persistent storage on the researcher-controlled application server; only nonaudio summaries and heuristic scores were retained locally.

\section{Results}

\begin{figure*}[!t]
\centering
\includegraphics[width=\textwidth]{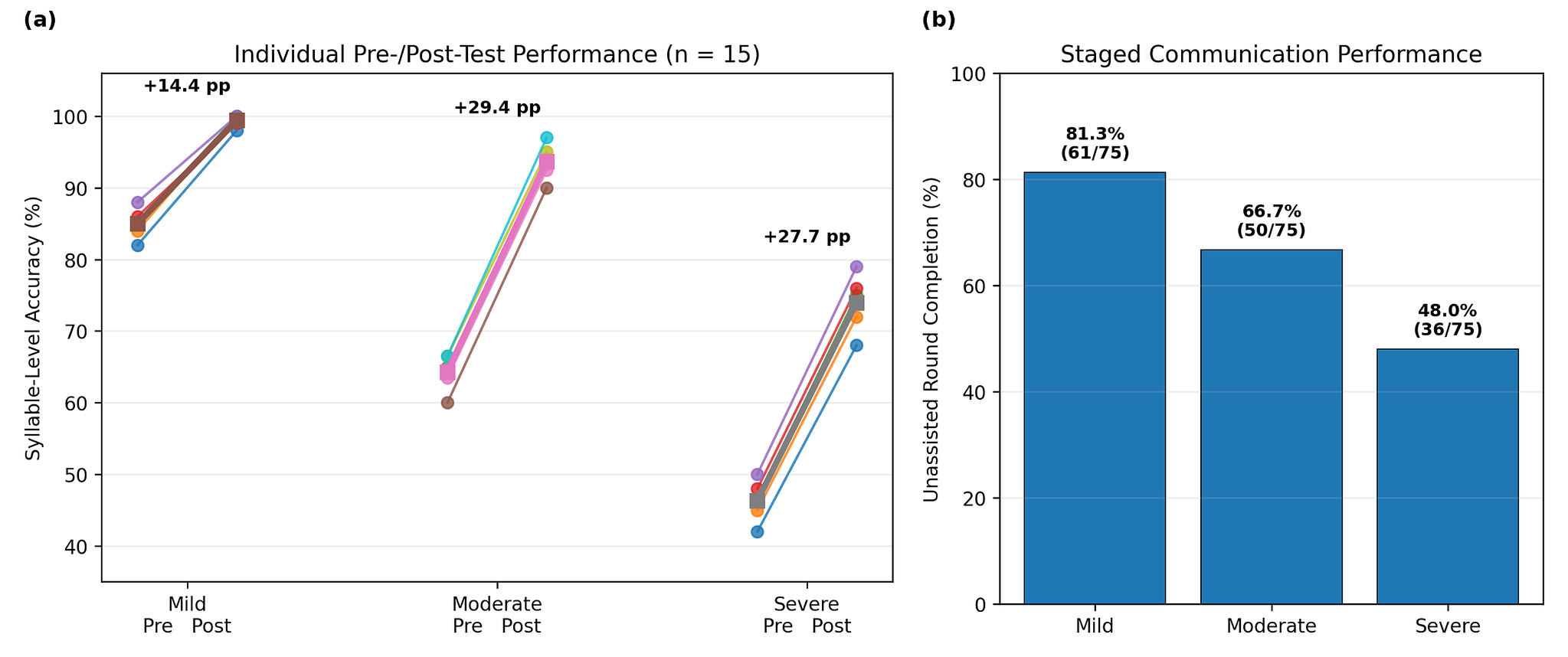}
\caption{Child pilot results. (a) Individual pre-/post-test syllable-level accuracy on matched parallel sentence sets. (b) Unassisted completion of staged communication rounds by severity group.}
\label{fig:pilot}
\end{figure*}

The individual pre/post trajectories and staged communication-completion rates are summarized in Fig.~\ref{fig:pilot}.

\subsection{Dataset Results}
Across the 13 dysarthric MDSC speakers, ClearTalk syllable-level accuracy was strongly associated with expert intelligibility ratings in the seed-42 analysis (Pearson $r = 0.908$, R-squared = 0.824, $p < 0.001$). The position-level analysis showed a descriptive decline in initial-consonant accuracy as expert-rated intelligibility decreased. The severe group had its lowest mean values for retroflex and alveolar/dental consonants, while control and mild speakers showed higher and more similar profiles. Because the MDSC severity groups were unbalanced (mild $n = 8$, moderate $n = 2$, severe $n = 3$), these position-level comparisons are interpreted descriptively rather than as confirmatory group tests.

\subsection{Expert Alignment and Robustness Results}
Table~\ref{tab:confusion} presents the target-consonant confusion matrix.

\begin{table}[!t]
\caption{Target-Consonant Confusion Matrix}
\label{tab:confusion}
\centering
\footnotesize
\setlength{\tabcolsep}{3.5pt}
\begin{tabular}{|l|c|c|c|}
\hline
 & \textbf{SLP Correct} & \textbf{SLP Error} & \textbf{SLP Deletion} \\
\hline
\textbf{System Correct} & 118 & 6 & 2 \\
\hline
\textbf{System Error} & 7 & 46 & 4 \\
\hline
\textbf{System Deletion} & 3 & 6 & 18 \\
\hline
\end{tabular}
\vspace{1pt}

\raggedright\footnotesize \textit{Note:} Overall agreement = 86.7\%; Cohen's kappa = 0.76; macro-F1 = 0.81.
\end{table}

Across the 15 children, the heuristic selected 39 of the 45 SLP-consensus targets, yielding a Top-3 overlap rate of 86.7 percent. The normalized-accuracy baseline matched 28 of 45 targets (62.2 percent), 24.5 percentage points lower than the heuristic result.

At the target-consonant level, system and SLP labels agreed on 182 of 210 cases (86.7 percent). Agreement beyond chance was substantial (kappa = 0.76, $p < 0.001$), with macro-F1 = 0.81; deletion was the most challenging class.

Across 58 expert-confirmed errors, recall was 79.3 percent overall, ranging from 90.0 percent for alveolar and 85.7 percent for velar targets to 69.2 percent for retroflex targets; the remaining positions ranged from 75.0 to 83.3 percent.

Across ten MDSC seeds, correlation remained stable (mean $r = 0.904$, standard deviation = 0.011, range 0.885--0.922). Across the $3 \times 3$ weight grid, 93.3 percent of baseline Top-3 targets were retained, indicating local parameter robustness.

\subsection{Child Articulation Results}
Mean syllable-level accuracy increased from 85.0 to 99.4 percent in the mild group, 64.3 to 93.7 percent in the moderate group, and 46.3 to 74.0 percent in the severe group, corresponding to gains of 14.4, 29.4, and 27.7 percentage points. Across all children, mean accuracy increased from 65.2 to 89.0 percent, corresponding to an increase of 23.8 percentage points; every participant improved. An exact paired Wilcoxon signed-rank test yielded $W = 0$, $p < 0.001$.

\subsection{Communication and Usability Results}
Independent communication-round completion rates were 81.3 percent (61/75), 66.7 percent (50/75), and 48.0 percent (36/75) for mild, moderate, and severe groups. Completion declined with severity, but nearly half of severe-group rounds were completed without adult prompting or physical assistance. Joint child--guardian ratings averaged 4.35/5; engagement was highest (4.6), followed by perceived pronunciation improvement (4.5), while voice-output quality remained above 4.0 (4.1).

\section{Discussion}
MDSC results show that ClearTalk's syllable-level metric captures broad between-speaker differences in expert-rated intelligibility. The heuristic aligned more closely with SLP consensus (86.7 percent) than normalized accuracy (62.2 percent), suggesting greater robustness to sparse observations. Although ASR-derived pinyin alignment is an operational proxy rather than narrow phonetic transcription, expert evaluation at two resolutions supported its use: Top-3 priorities overlapped 86.7 percent with SLP consensus, while target-consonant labels achieved 86.7 percent agreement, a kappa of 0.76, and a macro-F1 score of 0.81.

Target-consonant recall was highest for alveolar and velar errors and lowest for retroflex errors. The lower retroflex recall may reflect difficulty distinguishing acoustically overlapping fricative and affricate realizations, motivating confidence-aware alignment and direct phoneme-level expert comparison rather than stronger diagnostic claims.

The child study was designed as a balanced feasibility evaluation rather than a treatment trial. Gains occurred on matched post-test sentences not used during practice, and the within-child design reduced repetition and phonetic-composition confounds. Results are therefore interpreted as immediate transfer within the tested protocol; controlled repeated-session studies are needed for causal and durable effects. Despite lower completion with severity, the severe group independently completed 48.0 percent of semantic rounds.

\balance
\section{Conclusion}
This paper presented ClearTalk, a wearable system connecting interpretable Mandarin articulation analysis with personalized practice and staged communication. MDSC evaluation showed a strong, sampling-stable association with expert intelligibility. System-selected priorities achieved 86.7 percent overlap with blinded SLP consensus, and target-consonant predictions achieved 86.7 percent agreement, a kappa of 0.76, and a macro-F1 score of 0.81. The balanced child pilot showed immediate gains on matched, unpracticed materials and usable communication support across severity groups. These results support the feasibility of broad ASR-derived profiles for guiding wearable practice and everyday communication assistance.

\end{document}